\documentclass{article}
\usepackage{fullpage}
\usepackage[utf8]{inputenc}
\usepackage[T1]{fontenc}    % use 8-bit T1 fonts
\usepackage{hyperref}       % hyperlinks
\usepackage{url}
\usepackage{booktabs}       % professional-quality tables
\usepackage{amsfonts}       % blackboard math symbols
\usepackage{nicefrac}       % compact symbols for 1/2, etc.
\usepackage{microtype}      % microtypography
\usepackage{enumitem}
\usepackage{amsmath}
\usepackage{amssymb}
\usepackage{amsthm}
\usepackage{xcolor}
\usepackage{natbib}
\usepackage{graphicx}
\usepackage{subcaption}
\usepackage{hanging}
\usepackage{authblk}

\newcommand{\decision}{Y}
\newcommand{\papidx}{i}
\newcommand{\revidx}{j}
\newcommand{\revset}{\mathcal{R}}
\newcommand{\revcntr}{\revset^{\text{cntr}}}
\newcommand{\revtest}{\revset^{\text{test}}}
\newcommand{\stat}{\tau}
\newcommand{\size}{d}

\newcommand{\numpap}{m}
\newcommand{\numrev}{n}
\newcommand{\var}{v}
\newcommand{\indicator}{\mathbb{I}}
\newcommand{\meancntr}{M^{\text{cntr}}}
\newcommand{\meantest}{M^{\text{test}}}
\newcommand{\decisiontest}{X}
\newcommand{\cl}{A}
\newcommand{\openreview}{\texttt{openreview{}}}
\newcommand{\name}{resubmission}

\newcommand{\ICML}{ICML}
\newcommand{\ICMLyear}{ICML}
\newcommand{\commentedout}[1]{}

\title{Prior and Prejudice: The  {Novice Reviewers'} Bias against Resubmissions in Conference 
Peer Review}

\author[${}^\spadesuit$]{Ivan Stelmakh}
\author[${}^\spadesuit$]{Nihar B. Shah}
\author[${}^\spadesuit$]{Aarti Singh}
\author[${}^{\heartsuit \diamondsuit}$]{Hal Daum{\'e} III}

\affil[${}^\spadesuit$]{School of Computer Science, Carnegie Mellon University}
\affil[${}^{\heartsuit}$]{University of Maryland, College Park}
\affil[${}^{\diamondsuit}$]{Microsoft Research, New York}

\date{\vspace{-13pt} \texttt{\{stiv,nihars,aarti\}@cs.cmu.edu, hal@umiacs.umd.edu} 
}

\begin{document}

\maketitle

\begin{abstract}
    Modern machine learning and computer science conferences are experiencing a surge in the number of submissions that challenges the quality of peer review as the number of competent reviewers is growing at a much slower rate. To curb this trend and reduce the burden on reviewers, several conferences have started encouraging or even requiring authors to declare the previous submission history of their papers. Such initiatives have been met with skepticism among authors, who raise the concern about a potential bias in reviewers' recommendations induced by this information. In this work, we investigate whether reviewers exhibit a bias caused by the knowledge that the submission under review was previously rejected at a similar venue,  {focusing on a population of novice reviewers who constitute a large fraction of the reviewer pool in leading machine learning and computer science conferences}. We design and conduct a randomized controlled trial closely replicating the relevant components of the peer-review pipeline with $133$ reviewers  {(master's, junior PhD students, and recent graduates of top US universities)} writing reviews for $19$ papers. The analysis reveals that reviewers indeed become negatively biased when they receive a signal about paper being a resubmission, giving almost 1 point lower overall score on a  {10-point Likert item ($\Delta = -0.78,  \ 95\% \ \text{CI} = [-1.30, -0.24]$)} than reviewers who do not receive such a signal. Looking at specific criteria scores (originality, quality, clarity and significance), we observe that  {novice} reviewers tend to underrate quality the most.
\end{abstract}

\section{Introduction}

In contrast to many other fields of science, where journals are the only established venues for research publication, in machine learning (ML) and computer science (CS), conferences are considered to be equally or even more attractive~\citep{vrettas15venues}. While being as selective as top journals, leading conferences ensure much shorter turnaround time thereby facilitating timely research dissemination and allowing authors to quickly resubmit their work to the next conference if it gets rejected. However, the explosion in the number of submissions received by top conferences has considerably challenged the sustainability of the conference peer-review process since the number of qualified reviewers is growing at a much slower rate~\citep{sculley19tragedy, shah19principled}.

In an attempt to decrease the load on reviewers by discouraging resubmissions without substantial changes, several leading ML and CS conferences have started requesting or requiring authors to declare if a previous version of their submission was rejected at other peer-reviewed venues. For example, a top-tier conference in natural language processing EMNLP 2019 allowed authors to decide whether they want to disclose the past submission history and provide the summary of changes they made, making this information available only to senior committee members. A similar opportunity was offered at a leading conference in artificial intelligence and statistics (AISTATS 2017) with the exception that the information about past rejections was also available to regular reviewers. Additionally, the AISTATS 2017 conference implemented an automated review sharing with some past conferences, making these reviews visible to senior committee members after the initial reviewing was completed.

Other conferences make it mandatory for authors to disclose the past submission history: the NeurIPS conference --- one of the most popular conferences in ML, which in 2019 received more than six thousand submissions --- in 2020 required authors of previously rejected submissions to declare the changes they made to the current version of the paper. Another top conference in artificial intelligence (IJCAI 2020) went even further and made full reviews from past venues available to reviewers by requiring authors to include them in the submission file \emph{before} the actual paper. 

In addition to changes implemented by specific venues, the \href{https://openreview.net}{\openreview} platform --- a growing conference management system that hosts the leading deep-learning conference ICLR and other forums --- offers a novel approach towards managing conferences in a transparent manner by allowing organizers to make both accepted and rejected submissions  accompanied with full reviews publicly available. This option has been used by the ICLR conference since 2017 and all reviews for papers submitted to the conference since then are now publicly available, allowing subsequent reviewers (even at different venues) of rejected papers to consult them at any time.

All these steps are supposed to facilitate improving the quality of peer review as reviews from past venues may reduce the burden on reviewers by allowing them to ``quickly focus in on what previous issues were and how they may or may not have been fixed''~\citep{francis08thoughts, lin20neurips}. While being especially actual for the fields of ML and CS, similar discussion is also happening in other areas. For example, a survey conducted by~\citet{cals13editors} among editors of general medical journals showed that despite there being a concern that reviews from previous venues may bias subsequent reviewers, 45\% of the participating editors prefer authors to indicate whether a paper has been previously rejected (the survey does not indicate how this information is supposed to be used in the review process). Only 24\% of editors oppose this idea, and the rest are indecisive. 

Despite the interest of journal editors and conference program committees in reusing reviews from past iterations of the review cycle, authors are less enthusiastic: ICFP --- the ACM-sponsored conference on functional programming --- in 2020 removed the option to upload the past reviews since no authors took this advantage in the previous edition of the conference. This skepticism suggests that authors do not believe that revealing old reviews will increase the acceptance chances of their submission. 

Authors' concerns are in fact supported by a long line of research in psychology that establishes susceptibility of human judgements to various biases~\citep[see][for overview]{tversky74heuristics, kahneman2011thinking, gilovich02heuristics}, some of which are caused by additional (and sometimes irrelevant) information available to the decision-maker~\citep{baron88outcome, tversky74heuristics, fischhoff75hindsight, ross75perception, carretta83evidence}. Projecting this evidence on the peer-review context, we hypothesize that the knowledge that a paper was rejected at a previous venue may negatively bias reviewers' evaluations, leading to what in this study we call a ``\name{} bias''. In other words, reviewers may judge the paper differently depending on whether they are notified about this paper being a resubmission or not. 

To highlight the potential impact of the \name{} bias, we note that the peer-review process of the major machine learning conferences is far from being absolutely consistent and objective. Indeed, several controlled experiments~\citep{lawrence14neurips, price14neurips, pier18agreement} found very low degree of agreement between reviewers evaluating the same manuscript. This implies that the outcome of the review process heavily depends on the reviewers to whom the submission is assigned, introducing significant randomness in the final decisions. Hence, even a strong paper that deserves acceptance has a nontrivial chance of being rejected due to this randomness. The \name{} bias can \emph{amplify this unfairness by putting a previously rejected paper at disadvantage in the subsequent conferences}. Given that success in academia is largely determined by the publication profile of a researcher, the \name{} bias can have far reaching consequences not just for a particular paper, but more generally also for career trajectories of researchers due to the widespread prevalence of the Matthew effect (``rich get richer'') in academia~\citep{merton1968matthew, squazzoni2012saint}.

Therefore, in this work, we aim at testing for the presence of the \name{} bias in peer review. {Focusing on the population of novice reviewers}, in conjunction with a reviewer-recruiting process of \ICML{} 2020 (a top ML conference), we design and conduct a randomized controlled trial to test the following research hypothesis:

\begin{center}
\begin{minipage}{0.9\linewidth}
  \begin{hangparas}{8.8em}{1}
    \textit{Research Hypothesis:} The signal about rejection from the same or similar venue in the past, received by  {novice} reviewers before they read the paper, induces a bias in reviewers' evaluations. 
  \end{hangparas}
\end{minipage}
\end{center}

In our hypothesis, we do not specify the direction of the effect and note that while authors are concerned about a negative bias, the bias can hypothetically be positive. For example, a reviewer may think that previously rejected papers have gone through another iteration of revisions and improvements and might be better on average than papers that have not previously been peer reviewed. Therefore, in this work we also aim to confirm the direction of the effect (if it exists). 

 {Importantly, we caveat that in this study we target the population of novice reviewers and the findings we report in this paper must not be overgeneralized to the whole reviewer population. While the choice of the study participants is mostly justified by the difficulty of engaging senior reviewers in the experiment, we discuss below (Section~\ref{section:setup}) that novice reviewers constitute a large fraction of the leading ML and CS conferences reviewer pool.} 

In this work, we do not aim to support or undermine the idea of reusing the past reviews. Instead, the answer to our research question will inform conference organizers and journal editors about potential side-effects of reusing reviews, and will help them to carefully select the point during the peer-review pipeline (if at all) at which previous reviews become available to reviewers. For example, if the \name{} bias in peer review is present, then editors or program chairs may prefer to keep the past submission history hidden from reviewers in the initial stages of the process to ensure that the reviewers form an unbiased opinion about the paper first.  {Additionally, the results of our experiment can inform the current practices of novice reviewer training where more emphasis could be made on how to avoid the potential resubmission bias.}

Past research on human decision making decisively establishes the presence of various cognitive biases in human judgement that are relevant to our research hypothesis. For example, the famous anchoring effect~\citep{tversky74heuristics} manifests in human evaluations being dependent on an (irrelevant) piece of information received at the beginning of the decision task. We defer the discussion of the general literature on decision making and cognition to Section~\ref{section:discussion}, but underscore that these findings do not necessarily transfer to our setting. Indeed, from the perspective of the dual-process model of cognition~\citep{kahneman02representativeness, stanovich99rational, stanovich08independence}, biases are properties of an autonomous heuristic system (System 1) whereas the demanding and rational reviewing task invokes the analytic system (System 2) that can potentially override such biases.

%%%%%%%%%%%%%%%%%%%%%%%%%%%%%
\section{Related Literature} 

Our paper contributes to a long line of empirical works studying various aspects of the academic peer-review process and in this section we discuss the most relevant papers.

\citet{resnik20groupthink}, in the chapter of their book dedicated to peer review, discuss the presence of groupthink --- a strong desire of group consensus that results in all deviating ideas being rejected --- in peer review. One potential contribution to the overall groupthink effect is susceptibility of experts to social influence, as demonstrated by~\citet{teplitskiy19influence}. In their experiment, reviewers (faculty at US medical schools) first evaluated and rated submissions assigned to them, and then were exposed to scores presumably given to these submissions by other anonymous reviewers. Unbeknown to participants, these scores were randomly sampled to be either above or below their scores. The experiment demonstrated that 47\% of reviewers decided to update their scores, and in all but one case (out of approximately 185) the update was in the direction of the external scores. This finding indicates a strong impact of social influence on experts' evaluations. The experiment of~\citet{teplitskiy19influence} was conducted alongside a review process of grant applications in which real awards were distributed; in this setup, reviewers are generally expected to reach a consensus, hence, updates of review scores do not necessarily indicate a bias in reviewers' judgements of the proposal quality and instead can be seen as attempts to decrease the inconsistency of evaluations. In our work, we design the experiment to remove the group deliberation component of peer review by eliminating the discussion stage of the process, and measure the bias in reviewers' attitude towards the submission.

Another widely documented bias~\citep[][and others]{rosenthal79file,
moscati94pob, Callaham98pob, Emerson10pob} that manifests in peer review is the \textit{positive}-outcome bias also known as the \emph{file drawer problem}. This bias results in reviewers judging soundness of the experimental works depending on the outcome of the study: works in which the null hypothesis is rejected are rated more favourably than otherwise equal papers that show nonsignificant results. 

One difference between the social influence, file drawer problem and \name{} bias which may result in different cognitive heuristics responsible for the effects is the stage of the review process at which the stimulus is received by reviewers. In case of the social influence, reviewers get biased by signals they receive \emph{after} the review process and this bias can be attributed to the desire of consensus, attempt to improve the accuracy of a review or aim to achieve some other social goals~\citep{resnik20groupthink, teplitskiy19influence, cialdini04influence}. The file drawer problem is induced by information observed  \emph{during} reviewing and may be attributed to the desire of accepting ``newsworthy'' (i.e., positive) studies for publications~\citep{Callaham98pob, lynch07goodqual}. Finally, the incarnation of the \name{} bias we study in this work completes the picture and is related to the information reviewers receive \emph{prior} to reviewing submissions. 

Confirmatory bias --- a tendency of people to emphasize evidence that support their views and ignore or misinterpret those that do not --- is another relevant effect connected to prior beliefs of reviewers. It was identified in context of peer review by~\citet{mahoney1977publication} who conducted an experiment in which reviewers were exposed to different versions of the manuscript with identical experimental procedure but different directions of the obtained results (either confirming or disproving the beliefs of reviewers). It turned out that reviewers who received the version contradicting their theoretical perspective were significantly harsher in their evaluations than those who received the version supporting their views. In Section~\ref{section:discussion} we draw further connections between the \name{} and confirmatory biases. 

A separate line of work~\citep[][and others]{tomkins17wsdm, blank91effects, okike16sbvsdb} studies the presence of various biases, including gender, affiliation and fame biases in single-blind peer review (i.e., when author identities are visible to reviewers). The work of~\citet{tomkins17wsdm} is of particular interest as it identifies strong fame and affiliation biases\footnote{Meta-analysis they perform also reveals a bias against female-authored submissions.} in reviewers' evaluations. As a result of this work, WSDM --- a premier conference in web-inspired research --- switched to double-blind reviewing (i.e., author identities are hidden from reviewers), demonstrating the potential impact of empirical research on peer-review practices. 

 {Finally, our work contributes to the growing literature in computer science, both theoretical and empirical, that aims to understand and improve the conference peer-review process. These works develop methodologies to address various biases and other issues in peer review such as miscalibration~\citep{roos2011calibrate,ge13bias,wang2018your}, commensuration bias~\citep{noothigattu2018choosing}, strategic or dishonest behavior~\citep{aziz2019strategyproof,xu2018strategyproof, stelmakh2020catchme,jecmen2020manipulation}, biases with respect to author demographics~\citep{tomkins17wsdm,stelmakh2019testing,manzoor2020uncovering}, and methods for better assignments of reviewers to papers~\citep{Garg2010papers, charlin13tpms,kobren19localfairness,fiez2019super,stelmakh2018forall}. We envisage that the biases discussed in the current paper may be mitigated by a combination of policy guidelines and such computational techniques.}

%%%%%%%%%%%%%%%%%%%%%%%%%%%%%%%%%%%%
\section{Experimental Setup}
\label{section:setup}

To test our research hypothesis of presence of the \name{} bias, we design a randomized controlled experiment that replicates the relevant components of the peer-review pipeline of machine learning conferences while giving us more control over the resubmission signal received by reviewers. Two main components of the peer-review process are pools of papers and reviewers, and we now describe how these pools were constructed.

\medskip

\noindent \textbf{Papers.} We solicited $\numpap = 19$ anonymized preprints in various sub-areas of machine learning. To ensure that participants of the experiment cannot obtain a signal about a paper being a resubmission from anywhere outside of the experimental context, we restricted the pool of papers to works that had not yet been accepted to any conference or journal and had never been submitted to conferences hosted by the openreview platform or any other venue that makes the list of rejected submissions publicly available. We note that in machine learning it is common to publish  preprints on arXiv (\url{arxiv.org}) and we allow papers available on arXiv to be used in the experiment. Additionally, we allow papers that were previously presented at workshops --- less formal and prestigious venues without proceedings --- because it is also common to present a preliminary version of the work in a workshop and then submit the full version to the conference.

The final pool of papers consisted of working papers, papers under review, workshop publications and unpublished manuscripts. The papers were 6--12 pages long excluding references and appendices (a standard range for many ML and CS conferences) and were formatted in various popular journals' and conferences' templates with all explicit venue identifiers removed.

\medskip

\noindent \textbf{Reviewers.} The reviewer pool of a typical machine learning conference consists of researchers at various seniority levels, working in areas covered by the conference. Perhaps unique to ML, a recent surge in the number of papers submitted to leading conferences has forced organizers to expand the reviewer pool by relaxing the qualification bar, that is, by introducing rather junior researchers to the reviewer pool: for example, 33\% of the NeurIPS 2016 reviewers (1082 out of 3233) were graduate students~\citep{shah2017design}. Using data on the structure of the reviewer pool of the \ICMLyear{} 2020 conference, we estimate that approximately 35\% of the reviewers are rather junior individuals who self-nominated and satisfied the screening requirements of having at least two papers published in some top ML venues and being a reviewer for at least one top ML conference in the past. Overall, we conclude that a significant fraction of ML reviewers are novices and in this study we concentrate on the population of novice and junior reviewers. We admit that in this work we do not approximate the general reviewer population by leaving out more experienced reviewers, but notice that at the \ICML{} 2020 conference each submission was assigned to at least one reviewer from the aforementioned subset, making it a significant part of the reviewer community. 

To recruit participants, we messaged mailing lists of five large large, top US universities (CMU, MIT, UMD, UC Berkeley and Stanford) and targeted master's and junior PhD students working in ML-related fields. The invitation also propagated to a small number of students outside of these schools through the word of mouth. The recruiting materials contained an invitation to participate in the \ICML{} reviewer-selection experiment (selection of reviewers was indeed a key goal of the experiment and we studied the resubmission bias on top of it). Specifically, we asked participants to write a review for one paper and promised that those who provide a high-quality review will be invited to join the \ICMLyear{} 2020 reviewer pool. Being a reviewer at the flagship ML conference is a recognition of one's expertise and we expected that this potential benefit will motivate students to join our experiment. As a result, we received responses from $\numrev=200$ candidates, more than 90\% of whom were master's and PhD students or recent graduates of the aforementioned universities, and all of them were added to the pool of participants without further screening. The research hypothesis we study in this work may be sensitive to awareness of participants; therefore, we employed deception and did not reveal the dual goal of this work to participants, that is, subjects were unaware that in parallel with selecting reviewers for \ICML{}, we also want to measure the impact of the resubmission signal on the reviewers' attitude towards papers they review.

\medskip

The experimental procedure closely followed the initial stages of the standard ML conference peer-review pipeline and was hosted using Microsoft Conference Management Toolkit (\url{https://cmt3.research.microsoft.com}). First, we asked participants to express their interest in reviewing specific papers by entering bids that take the following values: ``Not Willing'', ``In a Pinch'', ``Willing''  and ``Eager''. Thirteen participants did not enter any bids and were removed from the pool. The remaining participants were active in bidding (mean number of ``Willing''  and ``Eager'' bids is 4.7) and we assigned all of them to 1 paper each, where we tried to satisfy reviewer bids, subject to a constraint that each paper is assigned to at least 8 reviewers. As a result, 186 participants were assigned to a paper they bid either ``Willing'' or ``Eager'' and 1 participant was assigned to a paper they bid ``In a Pinch'' (this participant did not bid ``Eager'' or ``Willing'' on any paper).

Finally, we instructed participants that they should review the paper as if it was submitted to the real \ICML{} conference with the exception that the relevance to \ICML{} and the formatting issues (e.g., page limit, margins) should not be considered as criteria. To help participants in writing their reviews, we provided reviewer guidelines adapted from NeurIPS instructions (see supplementary materials on the first author's website) that discuss the best practices of reviewing. We gave participants 15 days to complete the review and then extended the deadline for 16 more days to accommodate late reviews as our original deadline interfered with the final exams at various US universities and the US holiday period.

Unbeknown to participants, we allocated half of them to the test condition and half to the control condition uniformly at random. Participants in the control condition did not receive the resubmission signal while subjects in the test condition were notified that the paper they are reviewing was rejected at the NeurIPS 2019 conference. Since the goal of the experiment declared to participants was to test out a new approach towards recruiting reviewers, the presence of the resubmission signal could have been confusing for the participants as this information is irrelevant to the task. To ensure that the presence of the signal does not look odd to reviewers, we incorporated it in a small author checklist placed on the first page of each submission as shown in Figure~\ref{fig:checklist}. Participants in the control condition received submissions with a single-item checklist asking about the code submission, while participants in the test condition were additionally informed that the paper is a resubmission. 

{
\begin{figure}[t]
\centering
\begin{subfigure}{0.5\textwidth}%
  \centering
  \includegraphics[width=6.9cm]{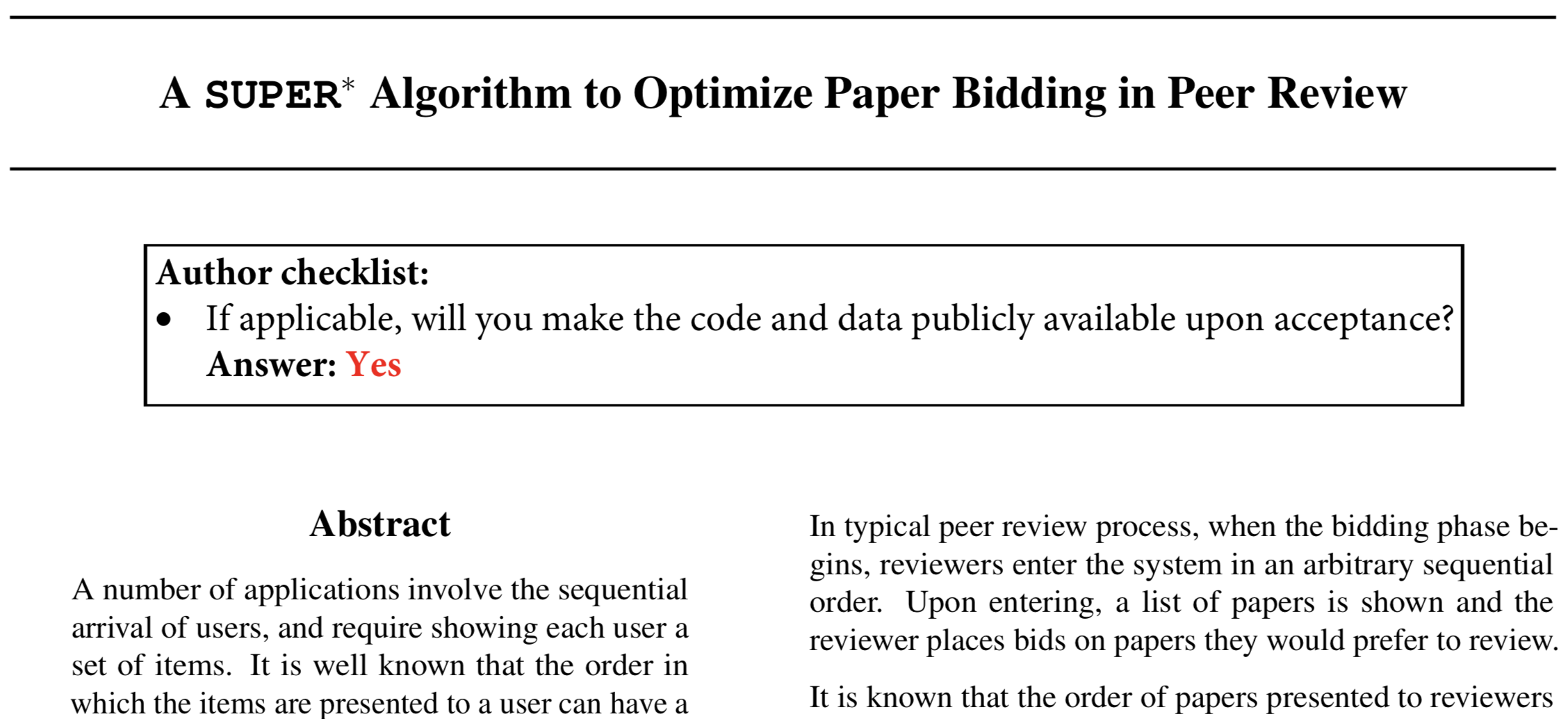}
  \caption{Control condition.}
  \label{fig:checklist_control}
\end{subfigure}%
\begin{subfigure}{0.5\textwidth}%
  \centering
  \includegraphics[width=6.9cm]{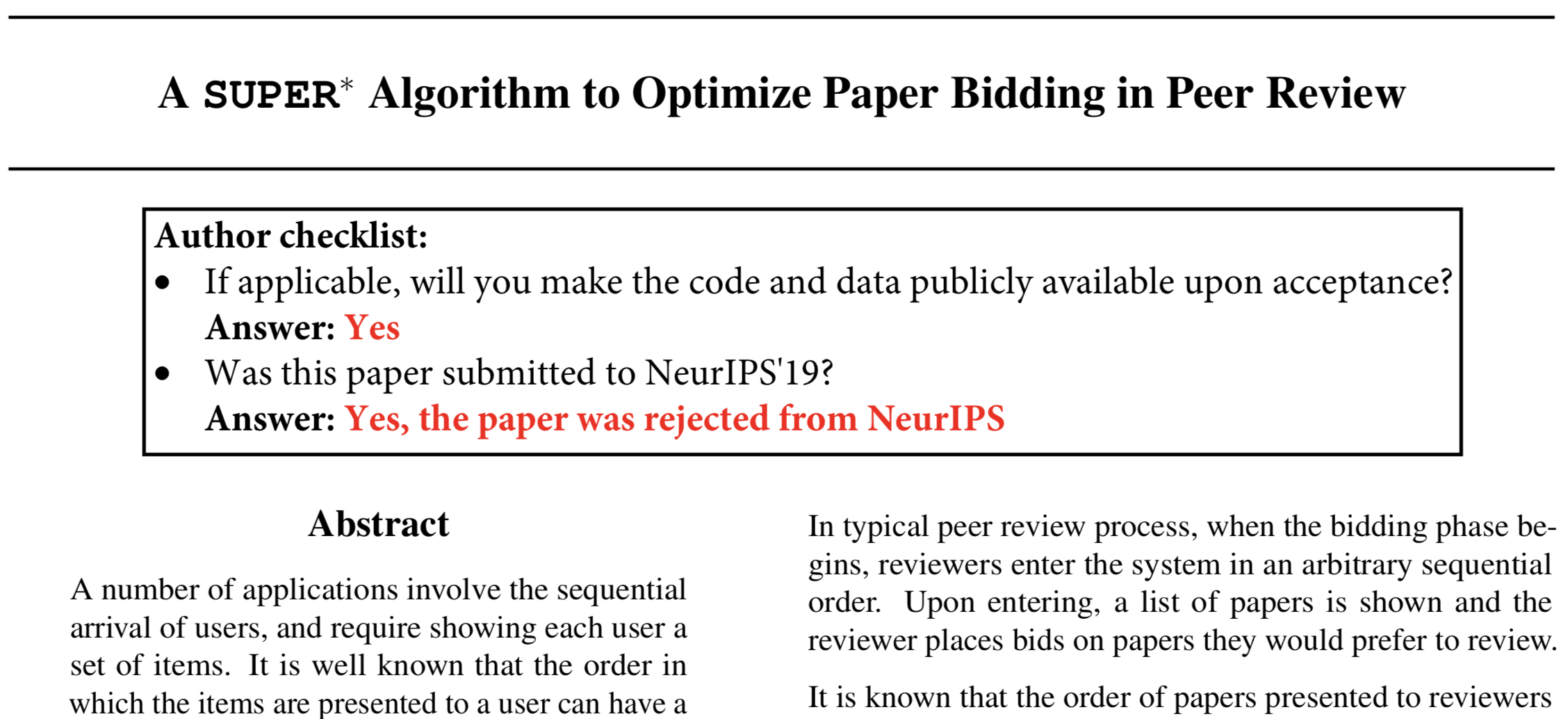}
  \caption{Test condition.}
  \label{fig:checklist_test}
\end{subfigure}%
\caption{Different versions of checklists shown to reviewers in the test and control conditions. The answer to all questions were ``Yes'' for all papers in both conditions.} 
\label{fig:checklist}
\end{figure}
}

%%%%%%%%%%%%%%%%%%%%%%%%%%%%%%%%%%%%
\section{Analysis of the Experiment}

Out of 187 participants who received a paper for review, 134 handed in the reviews (response rate of $71.7\%$). The remaining reviewers are excluded from subsequent analysis. Additionally, one of the participants misread the instructions and wrote a one-line review, rejecting the paper based on a minor violation of the standard \ICML{} page limit. This participant was also excluded from the analysis as formatting violation was not considered as a review criterion in the experiment. Table~\ref{table:aggregate} compares populations of reviewers between the test and control conditions using demographic information available to us. 

\begin{table}[b]
\vskip 0.15in
\begin{center}
\begin{small}
\begin{sc}
\begin{tabular}{lccc}
\toprule
          &\# Participants & \#With Prior Review Experience & \#With Publications \\
\midrule
Control 	& 68 	& 20 	& 47 	 \\ 
Test 	    & 65 	& 28 	& 49 	 \\ 
\bottomrule
\end{tabular}
\end{sc}
\end{small}
\end{center}
\caption{Comparison of demographics of reviewers across the test and control conditions. All differences are not significant at the level $\alpha = 0.1$.}
\label{table:aggregate}
\end{table}

The review form offered to participants (included in supplementary materials on the first author's website) was adapted from the NeurIPS 2019 form and contained 3 fields for open-ended feedback and 6 multiple choice questions in which reviewers were asked to give the overall score to the submission (10-point Likert item), evaluate the submission on 4 criteria (Originality, Quality, Clarity and Significance, 5-point Likert item for each criteria) and finally self-access their confidence in the scores assigned (5-point Likert item). Options offered for multiple choice questions were ordered from the most negative to the most positive and for the sake of analysis we associate these options to the numeric scales: 10-point scale for the overall score and 5-point scale for the other questions (larger numbers indicate more positive evaluations).\footnote{We release numeric evaluations reported by reviewers and the dataset is available on the first author's website.} In the sequel of this section, we compare these values between the test and control reviewers to see whether the resubmission signal significantly impacted the behaviour of reviewers in the test condition.

%%%%%%%%%%%%%%%%%%%%%%%%%%
\subsection{Testing procedure and effect size}

In this work, we employ a modification of the permutation test~\citep{fisher35permutation} that accounts for the fact that each paper is reviewed by different number of the test/control reviewers and is defined as follows. Recall that $\numpap$ stands for the number of papers in the experiment, for each paper $\papidx \in [\numpap]$ we let $\revcntr_{\papidx}$ (respectively $\revtest_{\papidx}$)  be a set of reviewers assigned to this paper in the control (respectively test) condition. Next, for any reviewer $\revidx \in \revcntr_{\papidx}$ we use $\decision_{\papidx}^{\revidx}$ to denote a numeric evaluation given by reviewer $\revidx$ to paper $\papidx$. Similarly, for any test reviewer $\revidx \in \revtest_{\papidx}$ we let $\decisiontest_{\papidx}^{\revidx}$ denote their evaluation of paper $\papidx$. With this notation, we define the test statistic:
\begin{align*}
    \stat = \frac{1}{\numpap} \sum\limits_{\papidx \in [\numpap]} \Bigg[ \underbrace{\frac{1}{|\revtest_{\papidx}|} \sum\limits_{j \in \revtest_{\papidx}} \decisiontest_{\papidx}^{\revidx}}_{\meantest_{\papidx}} - \underbrace{\frac{1}{|\revcntr_{\papidx}|} \sum\limits_{j \in \revcntr_{\papidx}} \decision_{\papidx}^{\revidx}}_{\meancntr_{\papidx}} \Bigg].
\end{align*}
Our test statistic compares mean scores $M_i^{\text{test}}$ and $M_i^{\text{cntr}}$ received by each paper $\papidx \in [\numpap]$ from the test and control reviewers, respectively, and then averages differences of these scores across submissions. A negative value of the test statistic implies that reviewers in the test condition are more harsh than those in the control condition, whereas a positive value indicates that the test reviewers are more lenient. 

Having defined the test statistic, we employ the permutation test to quantify the significance of its value. To this end, we permute reviewers within each paper between the test and control conditions uniformly at random, ensuring that the number of reviewers in each condition remains the same. We then recompute the value of the test statistic for 10,000 permutations and compare these values with the original value of the test statistic to obtain $P$-values.

In addition to reporting $P$-values of the test, we also provide the following measures of the effect size that capture different aspects of the effect:
\begin{itemize}[itemsep=5pt, leftmargin=15pt]
    \item  {\textbf{Simple Effect Size}  First, we note that our test statistic measures the between-conditions shift of the distributions of scores, that is, it represents the mean difference (in points of the corresponding Likert item) between scores given by two groups of reviewers. Hence, the value of the test statistic $\stat$ serves as a natural measure of the unscaled (simple) effect size~\citep{baguley08simple}. Note that the larger the absolute value of the test statistic, the stronger the effect.}
    
    \item  {\textbf{Scaled Effect Size (adaptation of Cohen's \textit{d})} Following recommendations of Cohen~\citep{cohen92power}, we supplement the aforementioned simple effect size (which is expressed in the original units of analysis) with its scaled version, making it independent of the corresponding units. While the scaling inevitably obscures the interpretation of the resulting value, it helps to compare effect sizes between evaluations of the overall score (measured on the 10-point Likert item) and criteria scores (measured on 5-point Likert items). To report the scaled effect size, we use an adaptation of the Cohen's $\size$~\citep{cohen92power}:} 
    \begin{align*}
        \size = \frac{\stat}{\sqrt{\var}},
\end{align*}
    where $\var$ is an upper bound on the variance of the terms $\meancntr_{\papidx}$ and $\meantest_{\papidx}$, $\papidx \in [\numpap]$, which is obtained by using the boundedness of numeric evaluations $\decision$ and $\decisiontest$. Formally, $\var = \nicefrac{(k-1)^2}{4}$, where $k$ is the number of points in the corresponding Likert item (10 for overall score and 5 for criteria scores). Note that we rely on the upper bound to avoid inflation of the effect size as terms corresponding to different papers have different variances, determined by the number of reviews written for this paper. The effect size computed in this manner is conservative because it assumes the largest possible sample variance. Similar to the simple effect size, the larger the absolute value of the scaled effect size, the stronger the effect. 

    \item \textbf{Relative Effect Size (stochastic superiority)}  To reduce the impact of the extreme numeric evaluations on the effect size, we also consider the measure of stochastic superiority that is computed taking into account relative rather than absolute differences~\citep{vargha00cl}:
    \begin{align*}
        \cl = \frac{1}{\sum\limits_{\papidx \in [\numpap]} |\revtest_{\papidx}| \times |\revcntr_{\papidx}|} \sum\limits_{\papidx \in [\numpap]} \left( \sum\limits_{\revidx_{1} \in \revtest_{\papidx}} \sum\limits_{\revidx_{2} \in \revcntr_{\papidx}} \left[ \indicator\left( \decisiontest_{\papidx}^{\revidx_1} > \decision_{\papidx}^{\revidx_2} \right) + 0.5 \cdot \indicator\left( \decisiontest_{\papidx}^{\revidx_1} = \decision_{\papidx}^{\revidx_2} \right) \right] \right),
    \end{align*}
    where $\indicator(\cdot)$ is an indicator function that equals 1 if its argument is correct and 0 otherwise. The denominator of this equation is the total number of (test, control) pairs of reviews written for the same paper. Each of these pairs contributes 1 to the numerator if the test review is more positive than the control review and 0.5 if the reviews contain the same numeric evaluation. Intuitively, this measure of the effect size equals to the empirical probability of a randomly sampled pair of (test, control) reviews written for the same paper having the test review more positive than the control (with ties broken uniformly at random). In contrast to the previous measures of the effect size, the further the value of $\cl$ from 0.5, the stronger the effect.
\end{itemize}

 {
In addition to reporting the point estimates of the effect sizes, we also report bootstrapped 95\% confidence intervals computed over 5,000 iterations, where bootstrapping is performed at the level of papers. With all the preliminaries introduced, we are now ready to present the results of our experiment.}

%%%%%%%%%%%%%%%%%%%%%%%%%%
\subsection{Results}

Table~\ref{table:joint_comparison} compares evaluations given by reviewers from the test and control conditions. The results indicate that reviewers in the test condition were significantly more strict in evaluating submissions, in average reporting almost 1 point lower overall score  {on the 10-point Likert item ($\Delta = -0.78,  \ 95\% \ \text{CI} = [-1.30, -0.24]$)} than reviewers in the control condition.  {The difference appears to be small but considerable and we provide more discussion of the strength of the effect in the next section.} Looking at the criteria score, we observe the similar trend of reviewers in the test conditions being stricter.  {Comparing unit-independent effect sizes (adaptation of Cohen's $\size$ and stochastic superiority), we note that the bias manifests the most in the evaluations of the submissions' quality which is known to be of high importance for the overall evaluation of the submission~\citep{noothigattu2018choosing}.}

\begin{table}[t!]
\vskip 0.15in
\begin{center}
\begin{small}
\begin{sc}
 {\begin{tabular}{lccccccr}
\toprule

          & & \multicolumn{2}{c}{Simple ES} & \multicolumn{2}{c}{Scaled ES} & \multicolumn{2}{c}{Relative ES}  \\
          \cmidrule(lr){3-4}
          \cmidrule(lr){5-6}
          \cmidrule(lr){7-8}
          & $P$-value & Size & 95\% CI & Size & 95\% CI & Size & 95\% CI \\
\midrule
Overall score  	& $.036$*  & $-0.78$ &  $[-1.30, -0.24]$	& $-0.17$ 	& 	$[-0.29, -0.05]$  & $0.42$ 	& $[0.32, 0.52]$  \\ 
Quality 	     	& $.005$*  & $-0.46$ & $[-0.69, -0.23]$& $-0.23$ 	& 	$[-0.35, -0.11]$  &$0.37$ 	&  $[0.27, 0.46]$ \\ 
Clarity 	    	& $.022$*  & $-0.44$ & $[-0.68, -0.19]$	    & $-0.22$ 	&  $[-0.34, -0.10]$    & $0.43$ 	& $[0.34, 0.52]$ \\ 
Significance 	 	& $.037$*  & $-0.36$ & $[-0.61, -0.10]$	& $-0.18$ 	& 	$[-0.30, -0.05]$  & $0.43$ 	& $[0.35, 0.50]$ \\ 
Originality 	 	& $.105$  & $-0.21$ & $[-0.40, -0.03]$	& $-0.11$ 	&  $[-0.20, -0.02]$  & $0.41$ 	&  $[0.32, 0.50]$ \\ 
Confidence  		& $.902$  & $-0.01$ & $[-0.20, \ \ \ 0.17]$	& $-0.01$ 	& 	 $[-0.10, \ \ \ 0.09]$   & $0.50$ 	&  $[0.42, 0.59]$ \\ 
\bottomrule
\end{tabular}}
\end{sc}
\end{small}
\end{center}

\caption{Comparison of evaluations given by participants in the test and control conditions to papers assigned to them for review. All $P$-values are two-sided and are computed by the permutation test with 10,000 permutations. Asterisk indicates significance at the level $\alpha = 0.05$.  {Confidence intervals for effect sizes are computed using 5,000 bootstrapped samples.}} 
\label{table:joint_comparison}
\end{table}

Overall, the data presented in Table~\ref{table:joint_comparison} supports our research hypothesis, suggesting that  {junior} reviewers are indeed susceptible to the \name{} bias. Notably, while the \name{} bias makes reviewers harsher, it does not seem to impact the confidence of reviewers. However, we qualify that self-evaluations of the confidence may not be a reliable measure of actual confidences, due to other biases manifesting in parallel. Indeed, as reviewers participate in the experiment to receive the invitation to join the pool of \ICML{} reviewers, they may be reluctant to report too high or too low confidence as it may hypothetically hurt their chances. This hypothesis agrees with the observed data as 122 out of 133 reviewers reported confidence level 3 or 4 on the 5-point Likert item (recall that larger values indicate higher confidence).

 {
As a final remark, we note that reported bootstrapped confidence intervals may be slightly more inaccurate than one would expect for the given sample size due to specific nature of data. Indeed, 133 subjects of the experiment were broken into two groups (test and control) and distributed across 19 papers. The bootstrapping was performed at the level of papers, that is, for each paper we bootstrapped test and control reviewers from the actual test and control reviewers assigned to this paper. The sample size for each paper is small and hence the resulting intervals could capture the excessive variance or underestimate the actual variance. Nevertheless, the combination of different measures of the effect size and the results of the permutation test (which is guaranteed to control the false alarm probability even under the small sample size) suggest that the effect is present in the data.} 

%%%%%%%%%%%%%%%%%%%%%%%%%%%%%%%%%%%%
\section{Discussion} 
\label{section:discussion}

The experiment we conduct in this work identifies the presence of the \name{} bias that manifests in  {junior} reviewers being significantly harsher in evaluating submissions that they know were previously rejected from similar venues. The design of our experiment ensures the absence of unobserved confounders that could drive the result, as we obtain reviews for the \emph{same} submission in both test and control conditions. In this section, we discuss some aspects of our experiment and suggest directions for future work.

\smallskip

\noindent \textbf{Strength of the effect} The size of the impact of the \name{} bias on the overall score received by submissions appears to be small according to the measures of the effect size we use in this work. We note, however, that top machine learning and computer science conferences are highly competitive and even small changes in reviewers' scores may have significant impact on the outcome of a submission and more generally on the researchers' career~\citep{thurner2011peer, squazzoni2012saint}. As a concrete example, data from the ICML 2012 conference~\citep{LangfordBlog} demonstrates that papers with mean reviewer score 2.67 (on a 4-point Likert item) were 6 times more likely to be accepted than papers with mean score 2.33: the difference between these scores is a single point decrease in a single review. While this data is observational and does not account for potential unobserved ``true quality'' of submissions, it suggests that even small effects may result in large changes in the outcomes.

Next, we note that reviewers' evaluations are known to be subjective and have a high variance~\citep{kerr1977manuscript,mahoney1977publication, ernst1994reviewer,bakanic1987manuscript,lamont2009professors}. As a result, the effect that would be perceived as ``strong'' in a hypothetical within-subject experiment may be much less prominent in the between-subject design due to additional variance in evaluations due to subjectivity.

Finally, the pool of papers we have is diverse as it contains both papers that are likely to be accepted to the top conferences and papers that are not. In absence of the ground-truth ranking of papers and due to a limited sample size, we cannot look closer at the strength of the bias as a function of a paper quality, but past work suggests that some biases may be especially prominent in a subset of borderline papers~\citep{blank91effects}. As a direction for future work, it would be interesting to understand the extent of the \name{} bias with a breakdown by the quality of the submission and some characteristics of reviewers (e.g., experience). 

\smallskip

\noindent \textbf{Population of participants} 
The data obtained in this experiment unfortunately does not allow us to decisively align the population of the experiment participants with the general population of top-tier machine learning conference reviewers. Moreover, most of participants of the experiment would not be invited to join the reviewer pool of the \ICML{} 2020 conferences through the standard ways of reviewer recruiting. However, as a result of the experiment, 52 participants whose reviews were found to be strong enough were invited to join the \ICMLyear{} reviewer pool. Therefore, these participants allow us to make some indirect comparisons to the general reviewer pool and we now present some relevant results. For a detailed analysis of performance of reviewers recruited through our experiment in the real \ICML{} conference, we refer the reader to the companion paper~\citep{stelmakh2020novice}.
 
 {
\begin{figure}[t]
\centering
\begin{subfigure}[t]{0.45\textwidth}%
  \centering
  \includegraphics[width=7cm]{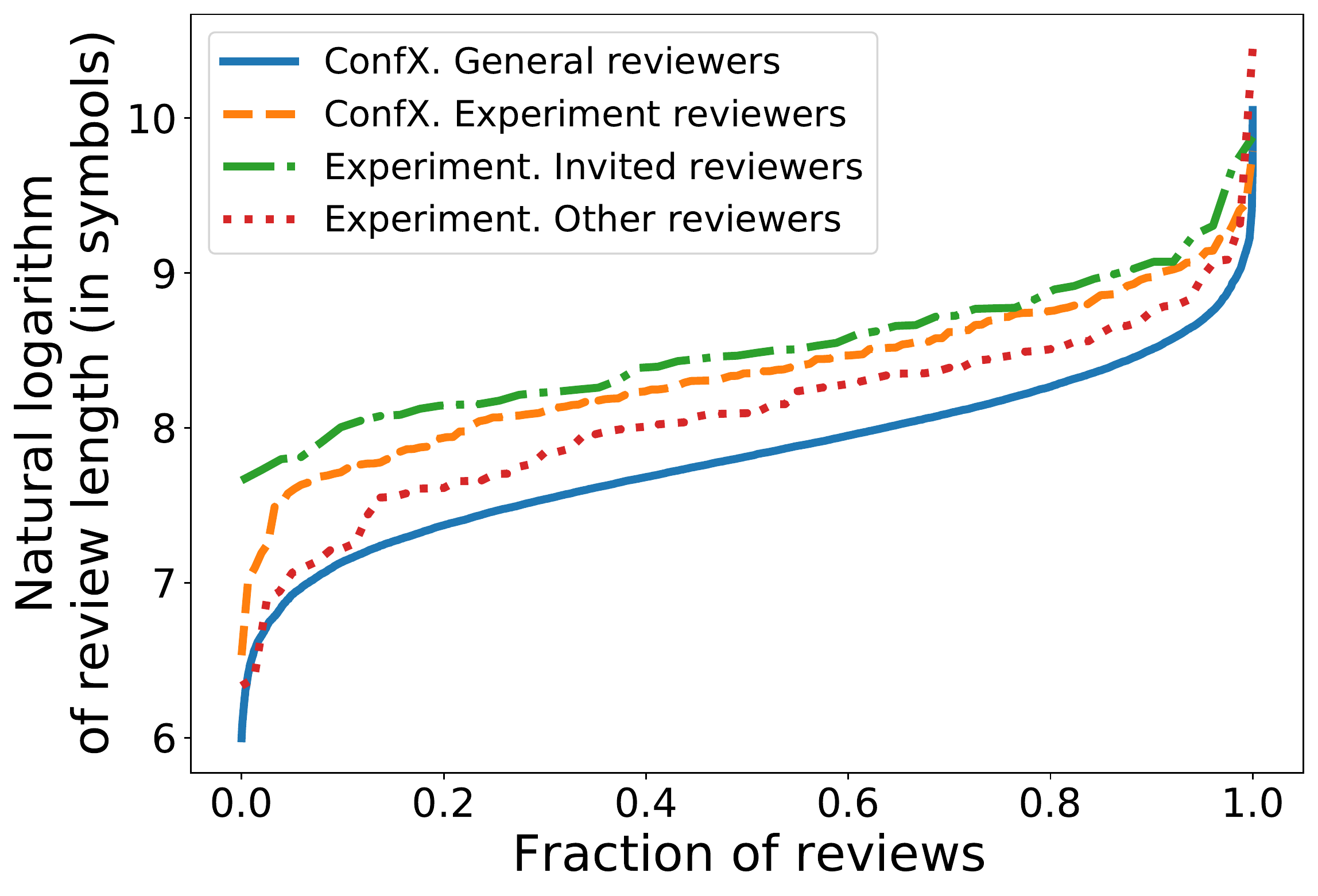}
  \caption{Distribution of the natural logarithm of the review length (in symbols). For clarity, we remove reviews shorter than 400 symbols from the set of reviews written by general \ICML{} reviews  {and plot linearly-interpolated curves instead of discrete points}.}
  \label{fig:loglencomp}
\end{subfigure}\qquad%
\begin{subfigure}[t]{0.45\textwidth}%
  \centering
  \includegraphics[width=7cm]{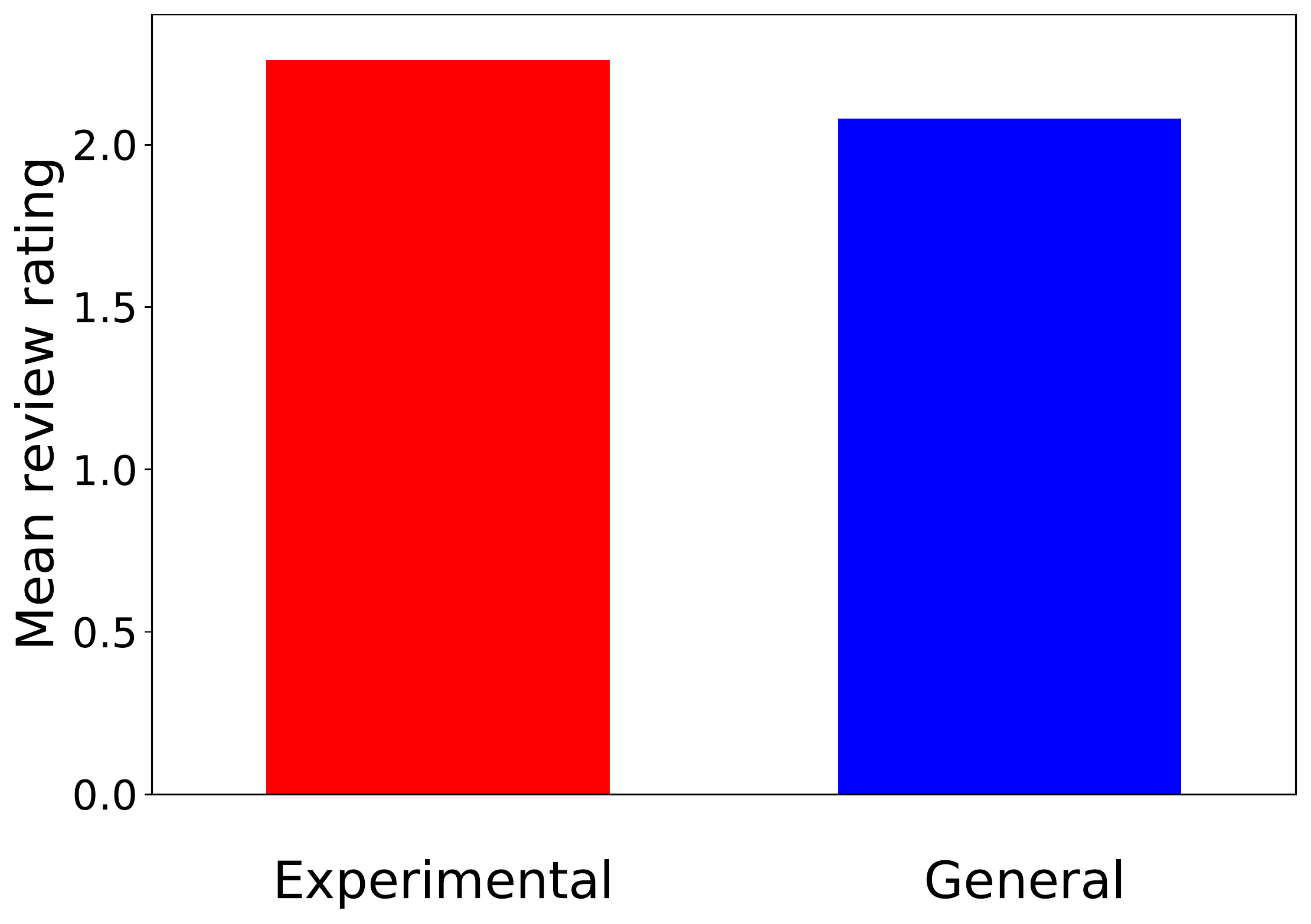}
  \caption{Mean rating of reviews given by area chairs across two groups of reviewers.}
  \label{fig:ratecomp}
\end{subfigure}%
\caption{Comparison of participants of the experiment with the general population of \ICML{} reviewers.}
\label{fig:comp}
\end{figure}
}

Figure~\ref{fig:loglencomp} juxtaposes the lengths of reviews written by different populations of reviewers at different venues. If we treat the length of a review as a measure of diligence, then we can see that reviewers selected in our experiment are not only more diligent than other participants of the experiment, but are also more diligent than the general reviewer population at the \ICML{} conference. Perhaps surprisingly, even if we consider the participants of our experiment who did not get invited to review for the main conference, the length of their reviews stochastically dominates the length of the reviews written by general \ICML{} reviewers with a significant margin, suggesting that they put a non-trivial effort in writing reviews. 

Next, Figure~\ref{fig:ratecomp} compares independent evaluations of review quality between the general population of \ICML{} reviewers and reviewers recruited through our experiment. In the \ICML{} review process each paper is assigned to several reviewers and one area chair. The area chairs are in charge of overseeing the reviewer activity, and at the end of the process, evaluate each reviewer on a 3-point Likert item: ``Failed to Meet Expectations'' (score 1), ``Met Expectations'' (score 2) and ``Exceeded Expectations'' (score 3). As seen in Figure~\ref{fig:ratecomp}, experimental reviewers appear to produce reviews of high quality according to the area chair's ratings, being on average better than general population of reviewers $( N_{\text{experimental}} = 111, \ N_\text{general} = 11624,\ \Delta = 0.18,\ P < .001)$.\footnote{To evaluate significance, we use a permutation test treating the rating of each review as an independent random variable.} 

As a word of caution, we note that comparisons we make in this section are based on observational data and may be influenced by confounding variables. For example, our experiment and the real conference employ different reviewer forms that could impact the length of the reviews. Additionally, experimental reviewers received at most 3 papers for review at \ICML{} and only 1 paper in the experiment while general \ICML{} reviewers received up to 6 submissions. There were also some other subtle differences between experimental and general reviewers in the assignment procedure of \ICML{} that could contribute to the observed result.

Subject to the aforementioned caveats, indirect evidence presented in Figure~\ref{fig:comp} suggests that the participants of the experiment who got invited to join the \ICML{} 2020 reviewer pool are comparable to the general population of top ML conference reviewers in terms of the length and quality of their reviews.

{However, we underscore again that the findings of this work must be interpreted with care and must not be overgeneralized to a more senior population of reviewers. Indeed, a past study of reviewers' behaviour~\citep{teplitskiy19influence} demonstrated that seniority and expertise of reviewers impact their behaviour and hence serve as confounding factors. Therefore, we do not know whether the results reported in this paper extend to the whole reviewer pool, and an interesting direction for future work is to understand the presence of the resubmission bias in the general population of reviewers.}

\smallskip

\noindent \textbf{Impact of this study on the reviewer-selection process} Recall that we conducted this study on top of the main experiment whose primary goal was to select participants to join the \ICML{} reviewer pool. To minimize the effect of this study on the main experiment, prior to the selection process we removed numeric evaluations given by participants from the reviews, so the selection was based solely on the textual part of reviews. We then manually analyzed reviews that fall in the study team members' area of expertise, asked authors to comment on the review quality, and crowdsourced expert opinions for reviews that we were unable to evaluate ourselves. Combining feedback from these sources, we eventually invited 52 reviewers to join the \ICML{} reviewer pool: 20 reviewers from the test condition and 32 reviewers from the control condition. The difference between the invitation rates appears to be insignificant at the level $\alpha = 0.05$ ($\Delta = 0.16, P = 0.075$).

\smallskip

\noindent \textbf{Cognitive mechanism of the \name{} bias} While in this work we do not aim to identify the cognitive mechanism of the \name{} bias, we now briefly comment on its relationship to several relevant cognitive biases known from general psychological literature.

\begin{itemize}[itemsep=2pt, leftmargin=15pt]
 
    \item \textit{Anchoring~\citep{tversky74heuristics, strack97anchoring, mussweiler01achoring}} Human judgements and evaluations are known to depend heavily on an initial piece of information (anchor), even when the anchor is obviously irrelevant (e.g., a number generated by the wheel of fortune). The signal about the previous outcome received by reviewers may be seen as an anchor that prevents them from adjusting to a positive opinion about the paper; therefore, anchoring may be responsible for the \name{} bias.
    
    \item \textit{Social proof~\citep{asch51conformity, baron96conformity, resnik20groupthink, cialdini04influence}} When being a part of a group, individuals are known to be susceptible to social influence, often exhibiting conformity to the opinion of the group. While in our experiment we removed the group deliberation component of the review process and participants were acting individually, they could consider the rejection from the previous venue as a decision made by a group of more experienced reviewers and decide to comply, deferring to the authority.
    
    \item \textit{Confirmatory bias~\citep{mahoney1977publication, Lord79BiasedAA, rabin99firstimpression, garcia20confirmatory}} Once having a belief about the state of the world, people tend to be selective in accepting new evidence: evidence that supports their views gets accepted more easily while contradicting evidence may be ignored or misinterpreted. In this paradigm, the \name{} bias can contribute to creating an initial belief that the paper under review is of a low quality that can later be exacerbated by a biased interpretation of strength and weaknesses of the submission.
    
    \item \textit{Hindsight bias~\citep{fischhoff75hindsight, hastie90hindsight, roese12hindsight} and Outcome bias~\citep{baron88outcome, allison96outcome, marshall93outcome, gino08noharm}} Hindsight bias (``I knew it all along'') is a tendency of people to overestimate the predictability of the event after it becomes observed. Outcome bias is a slightly different effect that manifests in distortion of human judgements of decisions. When a stochastic outcome of the decision is observed, even having all the information available to the decision-maker at the time of the decision, people tend to judge the quality of the decision differently depending on whether they view the outcome as good or bad. While these effects do not directly correspond to the \name{} bias we study in this work, they suggest that availability of outcome information can adversely impact the evaluations.
\end{itemize}
Despite all of the above biases are known to be present in human judgements, their presence in peer review is not obvious because of the nature of the task performed by reviewers. Indeed, the reviewing task is analytical and requires rational thinking, thereby potentially reducing the reliance on heuristics responsible for cognitive biases~\citep{stanovich99rational, kahneman02representativeness}. For example, \citet{gino08noharm}, in context of the outcome bias, show that the outcome information biases participants less when they use rational mindset. Given that the task conducted by participants of the experiment of~\citet{gino08noharm} is much less demanding than reviewing a paper, one could hypothesize that in peer review the cognitive heuristics may be overridden as reviewers naturally engage in rational and analytical thinking. However, our study identifies the presence of the \name{} bias in reviewers' judgements, demonstrating that the \name{} bias is strong enough to manifest even if reviewers put a non-trivial amount of cognitive effort into their work, as evidenced by the length of the reviews and expert judgements.

\smallskip

\noindent \textbf{On general idea of reusing reviews} In this work we do not aim to justify or oppose the idea of reusing reviews and do not say whether the \name{} bias is desirable or not. On the one hand, top-tier conferences are overloaded with substandard-quality papers making multiple rounds of submissions without major changes in a hope that they will get accepted at some point. Availability of past reviews could aid in identifying such submissions and optimizing the review efforts with respect to them, thereby reserving reviewers' effort for stronger papers. Additionally, when authors are aware of the \name{} bias, they may evaluate their work against higher standards and refrain from submitting works that are not yet in perfect shape to avoid the adverse effect on subsequent resubmissions, thereby further decreasing the load on reviewers.

On the other hand, the amount of randomness in the review process observed in several experiments~\citep{lawrence14neurips, price14neurips, pier18agreement} and in a multitude of anecdotal evidence (for example, see a survey of Nobel Prize laureates in economics by~\citealt{gans94mighty}) suggests that even strong papers that do not require revisions can be rejected by pure chance. It is also known~\citep{travis1991new,lamont2009professors} that works that are novel or not mainstream, particularly those interdisciplinary in nature, face significantly higher difficulty in gaining acceptance. All these works can be put at significant disadvantage if reviewers are exposed to the resubmission signal as they may not have major aspects to be improved.

While in this work we do not resolve the aforementioned dichotomy, we identify the presence of the \name{} bias in reviewers' decisions and measure its effect size, letting the community and conference organizers decide on its desirability. Specifically, to perform the randomized controlled trial we choose perhaps the simplest format of exposing the previous outcome to reviewers, omitting the content of reviews received by submissions at past venues and withholding author statements that could explain changes (if any) made in the manuscript after the previous rejection.

In its simplicity, our stimulus can perhaps be seen as a point on a spectrum between two ways of exposure to the resubmission signal employed in the real world: the openreview system allows subsequent reviewers to see that the paper was rejected and exposes any damning or hypercritical past reviews without letting authors present their view beyond what is stated in the discussion. In contrast, some other venues allow authors to present their case more clearly. Exploring the wider range of the aforementioned spectrum and identifying the desirable point on it (if any) is an interesting direction for future work.

Overall, the questions of whether the \name{} bias is desirable and how to reuse the reviews from the past venues in a fair and efficient way is an open question that requires a discussion in the community. In this work, we provide some concrete evidence that can help conference organizers to make informed decisions when designing the peer-review system. A promising direction for the future work is to evaluate the proposal of reusing reviews in a holistic manner, studying the interactions between different aforementioned positive and negative effects both theoretically and empirically.

\section*{Acknowledgments}

We thank authors of papers used in the experiment for contributing their works to this study. We are grateful to the support team of the Microsoft Conference Management Toolkit (CMT) who hosted our experiment and helped with many customization requests. We also thank authors of NeurIPS reviewer guidelines whose work we adapted to our setting.  This study was approved by Carnegie Mellon University Institutional Review Board.

This work was supported in part by NSF CAREER Award CIF: 1942124 and NSF CIF: 1763734.

\bibliographystyle{apalike}
\bibliography{references}

\begin{thebibliography}{}

\bibitem[Allison et~al., 1996]{allison96outcome}
Allison, S.~T., Mackie, D.~M., and Messick, D.~M. (1996).
\newblock Outcome biases in social perception: Implications for dispositional
  inference, attitude change, stereotyping, and social behavior.
\newblock In Zanna, M.~P., editor, {\em Advances in Experimental Social
  Psychology}, volume~28, pages 53 -- 93. Academic Press.

\bibitem[Asch, 1951]{asch51conformity}
Asch, S.~E. (1951).
\newblock Effects of group pressure upon the modification and distortion of
  judgments.
\newblock In Guetzkow, H., editor, {\em Groups, leadership and men; research in
  human relations}, page 177–190. Carnegie Press.

\bibitem[Aziz et~al., 2019]{aziz2019strategyproof}
Aziz, H., Lev, O., Mattei, N., Rosenschein, J.~S., and Walsh, T. (2019).
\newblock Strategyproof peer selection using randomization, partitioning, and
  apportionment.
\newblock {\em Artificial Intelligence}, 275:295--309.

\bibitem[Baguley, 2008]{baguley08simple}
Baguley, T. (2008).
\newblock Standardized or simple effect size: What should be reported?
\newblock {\em British journal of psychology (London, England : 1953)},
  100:603--17.

\bibitem[Bakanic et~al., 1987]{bakanic1987manuscript}
Bakanic, V., McPhail, C., and Simon, R.~J. (1987).
\newblock The manuscript review and decision-making process.
\newblock {\em American Sociological Review}, pages 631--642.

\bibitem[Baron and Hershey, 1988]{baron88outcome}
Baron, J. and Hershey, J.~C. (1988).
\newblock Outcome bias in decision evaluation.
\newblock {\em Journal of Personality and Social Psychology}, 54(4):569--579.

\bibitem[Baron et~al., 1996]{baron96conformity}
Baron, R.~S., Vandello, J.~A., and Brunsman, B. (1996).
\newblock The forgotten variable in conformity research: Impact of task
  importance on social influence.
\newblock {\em Journal of Personality and Social Psychology}, 71(5):915--927.

\bibitem[Blank, 1991]{blank91effects}
Blank, R.~M. (1991).
\newblock The effects of double-blind versus single-blind reviewing:
  Experimental evidence from the american economic review.
\newblock {\em American Economic Review}, 81(5):1041--1067.

\bibitem[Callaham et~al., 1998]{Callaham98pob}
Callaham, M.~L., Wears, R.~L., Weber, E.~J., Barton, C., and Young, G. (1998).
\newblock Positive-outcome bias and other limitations in the outcome of
  research abstracts submitted to a scientific meeting.
\newblock {\em JAMA}, 280(3):254--257.

\bibitem[Cals et~al., 2013]{cals13editors}
Cals, J.~W., Mallen, C.~D., Glynn, L.~G., and Kotz, D. (2013).
\newblock Should authors submit previous peer-review reports when submitting
  research papers? views of general medical journal editors.
\newblock {\em Annals of family medicine}, 11(2):179--181.

\bibitem[Carretta and Moreland, 1983]{carretta83evidence}
Carretta, T.~R. and Moreland, R.~L. (1983).
\newblock The direct and indirect effects of inadmissible evidence1.
\newblock {\em Journal of Applied Social Psychology}, 13(4):291--309.

\bibitem[Charlin and Zemel, 2013]{charlin13tpms}
Charlin, L. and Zemel, R.~S. (2013).
\newblock The {T}oronto {P}aper {M}atching {S}ystem: An automated
  paper-reviewer assignment system.

\bibitem[Cialdini and Goldstein, 2004]{cialdini04influence}
Cialdini, R. and Goldstein, N. (2004).
\newblock Social influence: Compliance and conformity.
\newblock {\em Annual review of psychology}, 55:591--621.

\bibitem[Cohen, 1992]{cohen92power}
Cohen, J. (1992).
\newblock A power primer.
\newblock {\em Psychological Bulletin}, 112(1):155--159.

\bibitem[Emerson et~al., 2010]{Emerson10pob}
Emerson, G., Warme, W., Wolf, F., Heckman, J., Brand, R., and Leopold, S.
  (2010).
\newblock Testing for the presence of positive-outcome bias in peer review: A
  randomized controlled trial.
\newblock {\em Archives of Internal Medicine}, 170(21):1934--1939.

\bibitem[Ernst and Resch, 1994]{ernst1994reviewer}
Ernst, E. and Resch, K.-L. (1994).
\newblock Reviewer bias: a blinded experimental study.
\newblock {\em The Journal of laboratory and clinical medicine},
  124(2):178--182.

\bibitem[Fiez et~al., 2020]{fiez2019super}
Fiez, T., Shah, N., and Ratliff, L. (2020).
\newblock A {SUPER}* algorithm to optimize paper bidding in peer review.
\newblock In {\em Conference on Uncertainty in Artificial Intelligence}.

\bibitem[Fischhoff and Beyth, 1975]{fischhoff75hindsight}
Fischhoff, B. and Beyth, R. (1975).
\newblock I knew it would happen: Remembered probabilities of once—future
  things.
\newblock {\em Organizational Behavior and Human Performance}, 13(1):1 -- 16.

\bibitem[Fisher, 1935]{fisher35permutation}
Fisher, R.~A. (1935).
\newblock {\em The design of experiments.}
\newblock Oliver \& Boyd, Oxford, England.

\bibitem[Francis, 2008]{francis08thoughts}
Francis, P. (2008).
\newblock Thoughts on improving review quality.
\newblock In {\em Proceedings of the Conference on Organizing Workshops,
  Conferences, and Symposia for Computer Systems}, WOWCS’08, USA. USENIX
  Association.

\bibitem[Gans and Shepherd, 1994]{gans94mighty}
Gans, J.~S. and Shepherd, G.~B. (1994).
\newblock How are the mighty fallen: Rejected classic articles by leading
  economists.
\newblock {\em Journal of Economic Perspectives}, 8(1):165--179.

\bibitem[Garcia et~al., 2020]{garcia20confirmatory}
Garcia, J.~A., Rodriguez-S{\'a}nchez, R., and Fdez-Valdivia, J. (2020).
\newblock Confirmatory bias in peer review.
\newblock {\em Scientometrics}, 123:517 -- 533.

\bibitem[Garg et~al., 2010]{Garg2010papers}
Garg, N., Kavitha, T., Kumar, A., Mehlhorn, K., and Mestre, J. (2010).
\newblock Assigning papers to referees.
\newblock {\em Algorithmica}, 58(1):119--136.

\bibitem[Ge et~al., 2013]{ge13bias}
Ge, H., Welling, M., and Ghahramani, Z. (2013).
\newblock A {B}ayesian model for calibrating conference review scores.

\bibitem[Gilovich et~al., 2002]{gilovich02heuristics}
Gilovich, T., Kahneman, D., and Gilovich, T. (2002).
\newblock Heuristics and biases: The psychology of intuitive judgment.
\newblock page 857.

\bibitem[Gino et~al., 2008]{gino08noharm}
Gino, F., Moore, D., and Bazerman, M. (2008).
\newblock No harm, no foul: The outcome bias in ethical judgments.
\newblock {\em Harvard Business School, Harvard Business School Working
  Papers}.

\bibitem[Hawkins and Hastie, 1990]{hastie90hindsight}
Hawkins, S.~A. and Hastie, R. (1990).
\newblock Hindsight: Biased judgment of past events after the outcomes are
  known.
\newblock {\em Psychological Bulletin}, 107(3):311--327.

\bibitem[Jecmen et~al., 2020]{jecmen2020manipulation}
Jecmen, S., Zhang, H., Liu, R., Shah, N.~B., Conitzer, V., and Fang, F. (2020).
\newblock Mitigating manipulation in peer review via randomized reviewer
  assignments.
\newblock In {\em NeurIPS}.

\bibitem[Kahneman, 2011]{kahneman2011thinking}
Kahneman, D. (2011).
\newblock {\em Thinking, fast and slow}.
\newblock Farrar, Straus and Giroux, New York.

\bibitem[Kahneman and Frederick, 2002]{kahneman02representativeness}
Kahneman, D. and Frederick, S. (2002).
\newblock Representativeness revisited: Attribute substitution in intuitive
  judgment.
\newblock {\em Heuristics and biases: The psychology of intuitive judgment},
  49:49–81.

\bibitem[Kerr et~al., 1977]{kerr1977manuscript}
Kerr, S., Tolliver, J., and Petree, D. (1977).
\newblock Manuscript characteristics which influence acceptance for management
  and social science journals.
\newblock {\em Academy of Management Journal}, 20(1):132--141.

\bibitem[Kobren et~al., 2019]{kobren19localfairness}
Kobren, A., Saha, B., and McCallum, A. (2019).
\newblock Paper matching with local fairness constraints.
\newblock In {\em ACM SIGKDD International Conference on Knowledge Discovery \&
  Data Mining}.

\bibitem[Lamont, 2009]{lamont2009professors}
Lamont, M. (2009).
\newblock {\em How professors think}.
\newblock Harvard University Press.

\bibitem[Langford, 2012]{LangfordBlog}
Langford, J. (2012).
\newblock {ICML} acceptance statistics.
\newblock \url{http://hunch.net/?p=2517} [Accessed: 05/30/2020].

\bibitem[Lawrence and Cortes, 2014]{lawrence14neurips}
Lawrence, N. and Cortes, C. (2014).
\newblock The {NIPS} experiment.
\newblock \url{http://inverseprobability.com/2014/12/16/the-nips-experiment}.
\newblock [Accessed: 05/30/2020].

\bibitem[Lin et~al., 2020]{lin20neurips}
Lin, H.-T., Balcan, M.-F., Hadsell, R., and Ranzato, M. (2020).
\newblock Getting started with {NeurIPS} 2020.
\newblock
  \url{https://medium.com/@NeurIPSConf/getting-started-with-neurips-2020-e350f9b39c28}.
\newblock Accessed: 05/30/2020.

\bibitem[Lord et~al., 1979]{Lord79BiasedAA}
Lord, C.~G., Ross, L.~D., and Lepper, M.~R. (1979).
\newblock Biased assimilation and attitude polarization : The effects of prior
  theories on subsequently considered evidence.
\newblock {\em Journal of Personality and Social Psychology},
  37(11):2098--2109.

\bibitem[Lynch et~al., 2007]{lynch07goodqual}
Lynch, J., Cunningham, M., Warme, W., Schaad, D., Wolf, F., and Leopold, S.
  (2007).
\newblock Commercially funded and united states-based research is more likely
  to be published; good-quality studies with negative outcomes are not.
\newblock {\em The Journal of bone and joint surgery. American volume},
  89:1010--8.

\bibitem[Mahoney, 1977]{mahoney1977publication}
Mahoney, M.~J. (1977).
\newblock Publication prejudices: An experimental study of confirmatory bias in
  the peer review system.
\newblock {\em Cognitive therapy and research}, 1(2):161--175.

\bibitem[Manzoor and Shah, 2020]{manzoor2020uncovering}
Manzoor, E. and Shah, N.~B. (2020).
\newblock Uncovering latent biases in text: Method and application to peer
  review.
\newblock In {\em INFORMS Workshop on Data Science}.

\bibitem[Marshall and Mowen, 1993]{marshall93outcome}
Marshall, G.~W. and Mowen, J.~C. (1993).
\newblock An experimental investigation of the outcome bias in salesperson
  performance evaluations.
\newblock {\em Journal of Personal Selling \& Sales Management}, 13(3):31--47.

\bibitem[Merton, 1968]{merton1968matthew}
Merton, R.~K. (1968).
\newblock The {M}atthew effect in science.
\newblock {\em Science}, 159:56--63.

\bibitem[Moscati et~al., 1994]{moscati94pob}
Moscati, R., Jehle, D., Ellis, D., Fiorello, A., and Landi, M. (1994).
\newblock Positive-outcome bias: Comparison of emergency medicine and general
  medicine literatures.
\newblock {\em Academic Emergency Medicine}, 1(3):267--271.

\bibitem[Mussweiler and Strack, 2001]{mussweiler01achoring}
Mussweiler, T. and Strack, F. (2001).
\newblock Considering the impossible: Explaining the effects of implausible
  anchors.
\newblock {\em Social Cognition - SOC COGNITION}, 19:145--160.

\bibitem[Noothigattu et~al., 2020]{noothigattu2018choosing}
Noothigattu, R., Shah, N.~B., and Procaccia, A.~D. (2020).
\newblock Loss functions, axioms, and peer review.
\newblock In {\em ICML workshop on Incentives in Machine Learning}.

\bibitem[Okike et~al., 2016]{okike16sbvsdb}
Okike, K., Hug, K.~T., Kocher, M.~S., and Leopold, S.~S. (2016).
\newblock Single-blind vs double-blind peer review in the setting of author
  prestige.
\newblock {\em JAMA}, 316(12):1315--1316.

\bibitem[Pier et~al., 2018]{pier18agreement}
Pier, E.~L., Brauer, M., Filut, A., Kaatz, A., Raclaw, J., Nathan, M.~J., Ford,
  C.~E., and Carnes, M. (2018).
\newblock Low agreement among reviewers evaluating the same nih grant
  applications.
\newblock {\em Proceedings of the National Academy of Sciences},
  115(12):2952--2957.

\bibitem[Price, 2014]{price14neurips}
Price, E. (2014).
\newblock The neurips experiment.
\newblock \url{http://blog.mrtz.org/2014/12/15/the-nips-experiment.html}.
\newblock [Accessed: 05/30/2020].

\bibitem[Rabin and Schrag, 1999]{rabin99firstimpression}
Rabin, M. and Schrag, J.~L. (1999).
\newblock First impressions matter: A model of confirmatory bias.
\newblock {\em The Quarterly Journal of Economics}, 114(1):37--82.

\bibitem[Resnik and Smith, 2020]{resnik20groupthink}
Resnik, D.~B. and Smith, E.~M. (2020).
\newblock {\em Bias and Groupthink in Science's Peer-Review System}, pages
  99--113.
\newblock Springer International Publishing, Cham.

\bibitem[Roese and Vohs, 2012]{roese12hindsight}
Roese, N.~J. and Vohs, K.~D. (2012).
\newblock Hindsight bias.
\newblock {\em Perspectives on Psychological Science}, 7(5):411--426.

\bibitem[Roos et~al., 2011]{roos2011calibrate}
Roos, M., Rothe, J., and Scheuermann, B. (2011).
\newblock How to calibrate the scores of biased reviewers by quadratic
  programming.
\newblock In {\em {AAAI} Conference on Artificial Intelligence}.

\bibitem[Rosenthal, 1979]{rosenthal79file}
Rosenthal, R. (1979).
\newblock The file drawer problem and tolerance for null results.
\newblock {\em Psychological Bulletin}, 86(3):638–641.

\bibitem[Ross et~al., 1975]{ross75perception}
Ross, L., Lepper, M., and Hubbard, M. (1975).
\newblock Perseverance in self-perception and social perception: Biased
  attributional processes in the debriefing paradigm.
\newblock {\em Journal of personality and social psychology}, 32:880--92.

\bibitem[Sculley et~al., 2019]{sculley19tragedy}
Sculley, D., Snoek, J., and Wiltschko, A.~B. (2019).
\newblock Avoiding a tragedy of the commons in the peer review process.
\newblock {\em CoRR}, abs/1901.06246.

\bibitem[Shah, 2019]{shah19principled}
Shah, N.~B. (2019).
\newblock Principled methods to improve peer review.

\bibitem[Shah et~al., 2018]{shah2017design}
Shah, N.~B., Tabibian, B., Muandet, K., Guyon, I., and Von~Luxburg, U. (2018).
\newblock Design and analysis of the {NIPS} 2016 review process.
\newblock {\em The Journal of Machine Learning Research}, 19(1):1913--1946.

\bibitem[Squazzoni and Gandelli, 2012]{squazzoni2012saint}
Squazzoni, F. and Gandelli, C. (2012).
\newblock Saint {M}atthew strikes again: An agent-based model of peer review
  and the scientific community structure.
\newblock {\em Journal of Informetrics}, 6(2):265--275.

\bibitem[Stanovich, 1999]{stanovich99rational}
Stanovich, K. (1999).
\newblock {\em Who Is Rational? Studies of Individual Differences in
  Reasoning}.
\newblock Mahwah, NJ: Erlbaum.

\bibitem[Stanovich and West, 2008]{stanovich08independence}
Stanovich, K. and West, R. (2008).
\newblock On the relative independence of thinking biases and cognitive
  ability.
\newblock {\em Journal of personality and social psychology}, 94:672--95.

\bibitem[Stelmakh et~al., 2018]{stelmakh2018forall}
Stelmakh, I., Shah, N., and Singh, A. (2018).
\newblock {PeerReview4All}: Fair and accurate reviewer assignment in peer
  review.
\newblock {\em arXiv preprint arxiv:1806.06237}.

\bibitem[Stelmakh et~al., 2019]{stelmakh2019testing}
Stelmakh, I., Shah, N., and Singh, A. (2019).
\newblock On testing for biases in peer review.
\newblock In {\em NeurIPS}.

\bibitem[Stelmakh et~al., 2020a]{stelmakh2020catchme}
Stelmakh, I., Shah, N.~B., and Singh, A. (2020a).
\newblock Catch me if i can: Detecting strategic behaviour in peer assessment.
\newblock In {\em ICML Workshop on Incentives in Machine Learning}.

\bibitem[Stelmakh et~al., 2020b]{stelmakh2020novice}
Stelmakh, I., Shah, N.~B., Singh, A., and {Daum{\'e} III}, H. (2020b).
\newblock A novice-reviewer experiment to address scarcity of qualified
  reviewers in large conferences.
\newblock Preprint
  \url{http://www.cs.cmu.edu/afs/cs.cmu.edu/user/istelmak/www/icml/novice.pdf}.

\bibitem[Strack and Mussweiler, 1997]{strack97anchoring}
Strack, F. and Mussweiler, T. (1997).
\newblock Explaining the enigmatic anchoring effect: Mechanisms of selective
  accessibility.
\newblock {\em Journal of Personality and Social Psychology}, 73:437--446.

\bibitem[Teplitskiy et~al., 2019]{teplitskiy19influence}
Teplitskiy, M., Ranub, H., Grayb, G.~S., Meniettid, M., Guinan, E.~C., and
  Lakhani, K.~R. (2019).
\newblock Social influence among experts: Field experimental evidence from peer
  review.

\bibitem[Thurner and Hanel, 2011]{thurner2011peer}
Thurner, S. and Hanel, R. (2011).
\newblock Peer-review in a world with rational scientists: Toward selection of
  the average.
\newblock {\em The European Physical Journal B}, 84(4):707--711.

\bibitem[Tomkins et~al., 2017]{tomkins17wsdm}
Tomkins, A., Zhang, M., and Heavlin, W.~D. (2017).
\newblock Reviewer bias in single- versus double-blind peer review.
\newblock {\em Proceedings of the National Academy of Sciences},
  114(48):12708--12713.

\bibitem[Travis and Collins, 1991]{travis1991new}
Travis, G. D.~L. and Collins, H.~M. (1991).
\newblock New light on old boys: Cognitive and institutional particularism in
  the peer review system.
\newblock {\em Science, Technology, \& Human Values}, 16(3):322--341.

\bibitem[Tversky and Kahneman, 1974]{tversky74heuristics}
Tversky, A. and Kahneman, D. (1974).
\newblock Judgment under uncertainty: Heuristics and biases.
\newblock {\em Science}, 185(4157):1124--1131.

\bibitem[Vargha and Delaney, 2000]{vargha00cl}
Vargha, A. and Delaney, H.~D. (2000).
\newblock A critique and improvement of the cl common language effect size
  statistics of mcgraw and wong.
\newblock {\em Journal of Educational and Behavioral Statistics},
  25(2):101--132.

\bibitem[Vrettas and Sanderson, 2015]{vrettas15venues}
Vrettas, G. and Sanderson, M. (2015).
\newblock Conferences versus journals in computer science.
\newblock {\em Journal of the Association for Information Science and
  Technology}, 66(12):2674--2684.

\bibitem[Wang and Shah, 2019]{wang2018your}
Wang, J. and Shah, N.~B. (2019).
\newblock Your 2 is my 1, your 3 is my 9: Handling arbitrary miscalibrations in
  ratings.
\newblock In {\em AAMAS}.

\bibitem[Xu et~al., 2019]{xu2018strategyproof}
Xu, Y., Zhao, H., Shi, X., and Shah, N. (2019).
\newblock On strategyproof conference review.
\newblock In {\em IJCAI}.

\end{thebibliography}

\end{document}